\documentclass[lettersize,journal]{IEEEtran}
\usepackage{amsmath,amsfonts}
\usepackage{algorithm}
\usepackage{amsmath, amssymb, amsfonts, bm, mathtools}

\usepackage{algpseudocode}
\usepackage{array}
\usepackage{xcolor}
\floatname{algorithm}{Algorithm}

\usepackage[caption=false,font=normalsize,labelfont=sf,textfont=sf]{subfig}
\usepackage{textcomp}
\usepackage{stfloats}
\usepackage{url}
\usepackage{verbatim}
\usepackage{graphicx}
\usepackage{cite}
\usepackage{multirow}

\usepackage{graphicx}

\begin{document}
%JH
%\title{Enhancing Fault Diagnosis through Statistically Balanced and Physically Coherent Synthetic Data: An F2GAN Approach}
\title{F2GAN: A Feature-Feedback Generative Framework for Reliable AI-Based Fault Diagnosis in Inverter-Dominated Microgrids}

\author{Swetha Rani Kasimalla,~\IEEEmembership{Graduate Student Member, IEEE}, Kuchan Park,~\IEEEmembership{Graduate Student Member, IEEE},\\ Junho Hong,~\IEEEmembership{Senior Member, IEEE}, and Young-Jin Kim,~\IEEEmembership{Senior Member, IEEE}
        % <-this % stops a space
\thanks{
	
		S.-R. Kasimalla, K. Park, J. Hong are with the Department of Electrical and Computer Engineering, University of Michigan -- Dearborn, Dearborn, MI, 48128 USA (emails: sweraka@umich.edu; kuchan@umich.edu; jhwr@umich.edu).}
\thanks{Y.-J. Kim is with the Department of
Electrical Engineering, Pohang University of Science and Technology, Pohang
37673, South Korea (e-mail: powersys@postech.ac.kr).
(corresponding author: Young-Jin Kim)}
	}

\markboth{IEEE TRANSACTIONS ON SMART GRID}%
{Shell \MakeLowercase{\textit{et al.}}: Enhancing Microgrid Protection Systems Through GAN- and GenAI-Generated Synthetic Fault Datasets}


\maketitle

\begin{abstract}
Enhancing the reliability of AI-based fault diagnosis in inverter-dominated microgrids requires diverse and statistically balanced fault datasets for training. However, the intrinsic scarcity and imbalance of high-fidelity fault data, particularly for rare inverter malfunctions and extreme external line faults pose a significant limitation to dependable model training and validation. To overcome this bottleneck, this paper introduces a unified framework that first models a detailed inverter-dominated microgrid and systematically generates multiple internal and external fault scenarios to mitigate class imbalance and data scarcity. Subsequently, an enhanced generative model called F2GAN (Feature Feedback GAN) is developed to synthesize high-dimensional tabular fault data with improved realism and statistical alignment. Unlike conventional GAN variants, F2GAN integrates multi-level feedback mechanisms based on mean–variance, correlation, and feature-matching loss functions, enabling the generator to iteratively refine output distributions toward real fault feature spaces.
The fidelity of the generated datasets is evaluated through quantitative and qualitative  analyses. To further assess the practical utility of synthetic data, Train-on-Synthetic, Test-on-Real (TSTR) experiments are conducted, demonstrating strong generalization of downstream machine learning classifiers trained exclusively on F2GAN-generated samples.
Additionally, the proposed framework is validated through a hardware-in-the-loop (HIL) fault diagnosis platform, which is integrated with a real-time simulator and a graphical user interface (GUI). This system enables real-time monitoring and classification of fault type and location using a neural network model trained on synthetic data and the real simulated data, achieving 100$\%$ diagnostic accuracy under real-time testing conditions. The combined results confirm that F2GAN effectively bridges the data gap between simulated and real-world microgrid fault datasets.

\end{abstract}

\begin{IEEEkeywords}
Microgrids, fault diagnosis, synthetic data generation, generative adversarial networks (GAN), feature feedback, class imbalance, HIL validation.
\end{IEEEkeywords}

\section{Introduction}
The growing penetration of inverter-based distributed generators (IBDGs) within microgrids has transformed modern distribution systems into highly dynamic, inverter-dominated networks. While this transition enhances flexibility, reliability, and the integration of renewable energy sources, it simultaneously introduces several challenges in fault diagnosis. In such systems, the traditional assumption of unidirectional fault current flow no longer holds, as distributed generation alters network topology and bidirectional fault currents emerge \cite{7467562,6308748}. Furthermore, inverter controllers inherently limit short-circuit currents (typically 1.5--2.0 p.u.), which complicates the sensitivity of overcurrent-based protection, especially during islanded operation. This limitation leads to rapid voltage and frequency deviations, often causing distributed units to disconnect due to out-of-zone faults before the main protection can act. 

Beyond external line faults, inverter-level internal faults represent another major concern, since semiconductor and converter failures account for nearly 80\% of inverter outages in microgrids \cite{10563400}. As inverter-based systems become increasingly critical, the demand for intelligent fault diagnosis methods that can distinguish complex fault modes has grown significantly. Machine learning (ML)-based fault classification has shown strong potential for both external and internal fault detection \cite{9780384, 10416843,10116186, 11204575} due to its ability to learn non-linear relationships between system features. However, such models are inherently data-dependent, requiring sufficient fault event diversity and balanced class representation to achieve generalization in real operating conditions.

Unfortunately, obtaining large-scale, balanced, and diverse fault datasets remains a pressing challenge. Fault experiments are limited by safety, equipment stress, and privacy constraints in real utility systems. As a result, most available data are scarce, imbalanced, and largely statistical rather than temporal, because fault diagnosis is mainly based on snapshot data. Multiple snapshots are used to train the models, which help analyze the behaviors of the faults. Therefore to alleviate data scarcity, recent studies have explored generative models such as Conditional GANs (CGAN) \cite{10818412}, Wasserstein GANs (WGAN-GP) \cite{9854145}, Variational Autoencoder-based GANs (VAE-GAN) \cite{10171153}, and Temporal VAEs (TVAE) for various energy applications including load forecasting \cite{9731506}, renewable scenario generation \cite{9765318}, and transient stability analysis \cite{10681285}. 
For instance, an instability sample generation framework using GANs was proposed in \cite{10681285} for online transient stability assessment (TSA), 
where a GAN-based model was trained on temporal dynamic trajectories to balance the long-tailed distribution of instability cases. 
Similarly, an unbalanced graph-GAN architecture (UG-GAN) was developed in \cite{9911768} to synthetically generate three-phase unbalanced distribution system connectivity, 
capturing structural relationships through random-walk representations of network graphs rather than numeric fault data. 
Both works demonstrate the promise of GANs in generating rare or unobservable operating scenarios; however, their focus remains on system-level or graph-based temporal data, rather than physical feature-level fault patterns. 

In the renewable energy domain, several studies have employed GAN-based architectures for time-series prediction and scenario generation. 
For example, \cite{10609361} utilized a time-series GAN for wind power output correction under extreme weather conditions, while \cite{9765318} introduced a style-based scenario generator coupled with a sequence encoder to reproduce spatial–temporal renewable dynamics. 
These approaches effectively capture temporal dependencies but lack mechanisms to ensure per-feature statistical coherence, which is crucial when dealing with voltage–current fault signatures. 
A transfer-learning-integrated improved GAN was proposed in \cite{10430431} to augment operation data under data-scarce conditions, demonstrating improved generalization with limited samples, yet without validation under dynamic or real-time conditions. 
More recently, diffusion-based generative models such as EnergyDiff \cite{11049025} have been applied to high-resolution time-series generation in energy systems, emphasizing temporal consistency but still neglecting multi-feature physical dependencies.

Generative approaches have also been applied to extreme event modeling and fault data augmentation. 
An auxiliary classifier GAN (ACGAN) was presented in \cite{10818412} to address data imbalance in fault detection and classification tasks, with a focus on improving the representation of minority classes. 
Likewise, the improved WGAN-GP model in \cite{9854145} generated synthetic samples for partial-discharge source classification, enabling stable training and improved performance on small datasets. 
While both works demonstrate the effectiveness of GANs for mitigating data imbalance, they primarily target single-domain classification tasks and do not incorporate statistical feedback mechanisms to preserve inter-feature relationships. 
Similarly, \cite{10208147} proposed MultiLoad-GAN to generate realistic load profiles for different consumers, and \cite{9731506} presented a super-resolution GAN (ProfileSR-GAN) for upsampling low-resolution load profiles, both confirming the utility of GANs for time-series load data. However, these approaches are limited to forecasting or resolution enhancement rather than diagnostic data synthesis.

A hybrid VAE-GAN architecture was introduced in \cite{10171153} to generate energy consumption time-series in smart homes, achieving diverse yet coherent load patterns. 
Although such hybrid models bridge generative diversity with distribution learning, they are inherently designed for continuous temporal data, rather than high-dimensional tabular datasets like fault snapshots in microgrids. While these prior works leverage GANs for time-series augmentation, network topology synthesis, or renewable forecasting, 
they predominantly operate in offline, statistical, or temporal domains and need to be tested on snapshot datasets. Crucially, not all GAN architectures are universally applicable across all types of data. The structure and design of a GAN must be tailored to the specific characteristics of the dataset and application. As highlighted by  Chundawat et al.,\cite{9984938} structural GANs require custom architecture and validation strategies that depend on the type of dataset (e.g., tabular, graph-based, or image-like). These considerations are especially important for microgrid fault diagnosis, where maintaining inter-feature physical dependencies such as voltage-current relationships are critical for reliability and interpretability.

%Hence this paper focuses on addressing the synthesis of physically coherent, feature-consistent fault data for ML-based fault diagnosis, and validate this synthetic data effectiveness under real-time temporal environments with HIL validation. Hence this paper introduces the F2GAN, a novel GAN architecture designed to embed feedback of statistical and physical consistency during training. Unlike conventional GANs, F2GAN incorporates feedback terms for mean, variance, and correlation between generated and real features—guiding the generator to learn not only data distributions but also inter-feature behavioral patterns. This allows the synthetic fault data to retain statistical fidelity while preserving the physical relationships among electrical quantities such as voltage, current. The proposed method enables class-balanced data generation for 12 internal and 30 external fault categories in a microgrid testbed. The synthesized data are subsequently used to train ML-based fault classifiers, which are validated on HIL testbed enabled with GUI for fault monitoring. The experimental results show that models trained on F2GAN-generated data achieve superior performance and robustness in temporal testing conditions, demonstrating that statistically synthesized data can effectively generalize to dynamic, real-world environments. Therefore the main contributions of this paper are summarized as follows:
Hence this paper presents a novel framework for synthesizing statistically and physically consistent snapshot-based fault data for microgrid fault diagnosis using an enhanced generative model. Unlike conventional GANs, F2GAN integrates statistical feedback mechanisms specifically mean, variance, and inter-feature correlation directly into the generator’s loss function. This guides the model to capture inherent voltage, current relationships across diverse fault types, enabling the generation of realistic and class-balanced datasets for 12 internal and 30 external fault scenarios. The synthetic data, comprising multiple high-fidelity fault snapshots, is used to train machine learning classifiers, which are deployed and validated on the HIL testbed equipped with a Python-based GUI for fault detection and localization. Experimental results confirm that F2GAN-trained models maintain high accuracy under dynamic operating conditions. The main contributions of this paper are summarized below
\begin{itemize}
    \item A statistically regularized GAN framework is developed, incorporating feedback metrics of mean, variance, and feature-wise correlation within the generator’s backpropagation. This formulation enforces both statistical and physical consistency, facilitating the coherent generation of realistic fault snapshot data.
    
    \item  F2GAN effectively mitigates data scarcity and class imbalance by generating diverse, physically consistent samples for a broad range of internal and external fault scenarios, while preserving essential voltage–current dependencies. 
    
    \item A  benchmarking analysis is carried out against state-of-the-art generative models tailored for tabular data, including CGAN, WGAN-GP, and TVAE. The comparative assessment encompasses both qualitative and quantitative analysis, wherein the TSTR accuracy is used to evaluate the generative fidelity and distributional alignment under varying model configurations.
    
    \item The synthetic datasets produced by F2GAN are utilized to train machine learning classifiers, which are subsequently deployed on the HIL testbed integrated with a Python-based GUI dashboard. This setup enables fault type and location diagnosis, thereby validating the proposed framework’s robustness and practical efficacy under real-time operational conditions.
\end{itemize}

The remainder of this paper is organized as follows.
Section~\ref{sub:Methodology} details the methodology of proposed F2GAN architecture and microgrid simulation framework.
Section~\ref{sub:Evaluation Metrics} describes the evaluation metrics and experimental setup. 
Section~\ref{sec:Results} presents the results of synthetic data evaluation and real-time validation, and 
finally, Section~\ref{sec:Conclusion} concludes the paper.

\section { Methodology }
\label{sub:Methodology}
%%%%%%%%%%% A %%%%%%%%%%%%%%
\subsection{Integrated Framework for Data Augmentation and Quality
Assessment}

\begin{figure*}[!t]
    \centering
    \includegraphics[width=\textwidth]{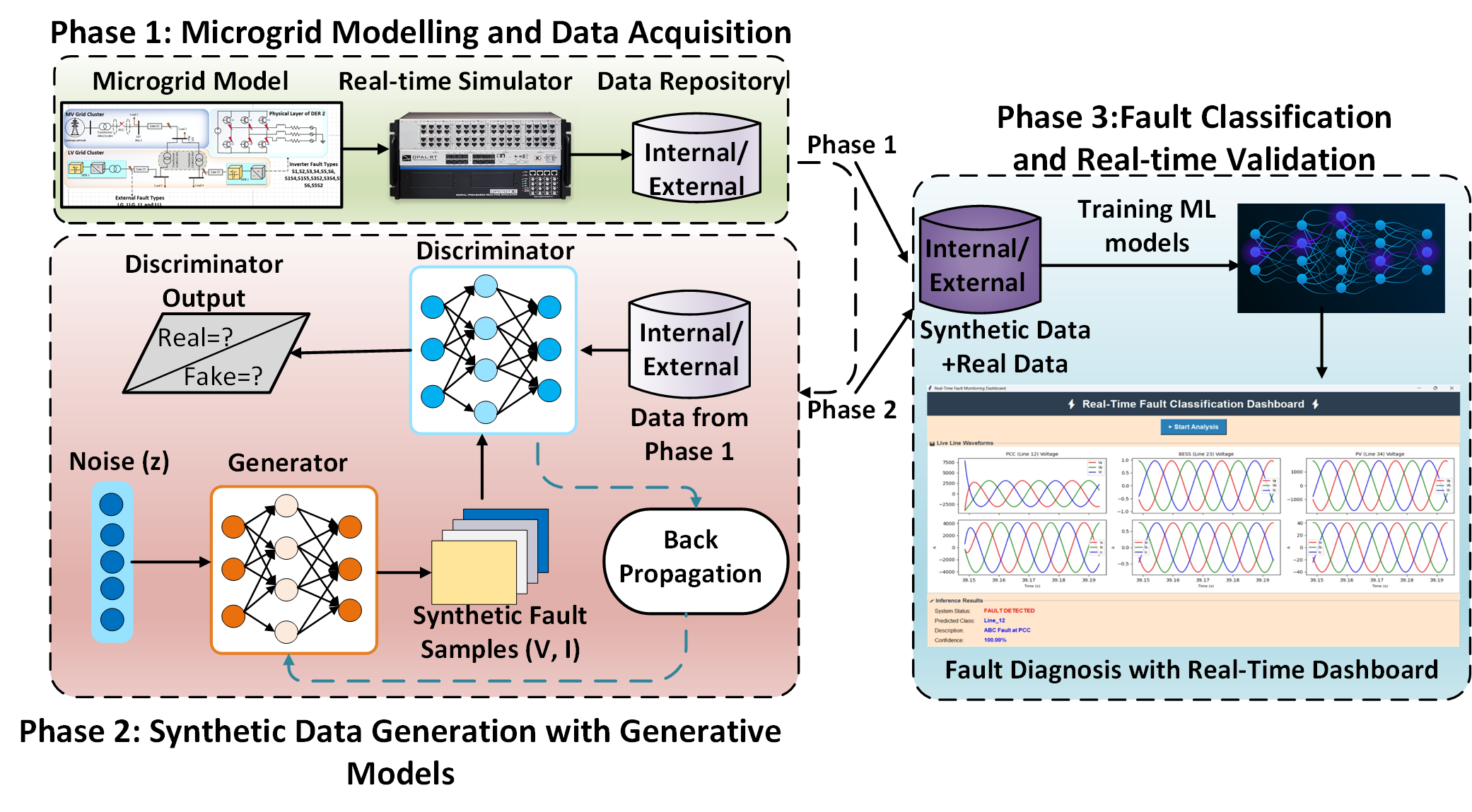} % Use \textwidth for full width
    \caption{Architecture for Fault Data Generation and Detection: This framework illustrates the generation of simulation data from a real-time simulator (Phase 1), the traditional GAN-based architecture for synthesizing internal and external fault data (Phase 2), and the training of a fault detection model to classify fault scenarios accurately (Phase 3).}
    \label{Fig1}
  
\end{figure*}
The overall architecture of the proposed fault data augmentation and diagnosis framework is illustrated in Fig.~\ref{Fig1}. The pipeline is structured into three sequential phases, each designed to address a critical component of the synthetic data generation and real-time validation workflow:
\begin{itemize}
    \item \textit{Microgrid Modeling and Data Acquisition (Phase 1):}
This phase establishes the foundational dataset for training and validation which is considered as real data. A microgrid model comprising internal and external fault scenarios is developed and deployed on a real-time HIL testbed. Voltage and current root mean square (RMS) snapshots are collected under various fault conditions to construct two distinct datasets: one for internal faults and another for external faults. The modeling and data acquisition methodology are elaborated in Section~\ref{subsec:microgrid_model}.
\item \textit{Synthetic Data Generation with Generative Models (Phase 2):}
In this stage, both datasets undergo augmentation using generative models. As illustrated in Fig.~\ref{Fig1}, Phase 2 outlines the architecture of a baseline GAN model, while the enhanced F2GAN structure is detailed in Fig.~\ref{Fig3}. The proposed F2GAN incorporates statistical feedback mechanisms to improve feature-level consistency. This phase is critical in addressing challenges related to class imbalance and data scarcity by synthesizing statistically and physically consistent fault snapshots. The augmentation strategies for both baseline and proposed models are comprehensively discussed in Sections~\ref{subsec:CGAN} to \ref{subsec:F2gan}.
\item \textit{Fault Classification and Real-Time Validation (Phase 3):}
The augmented datasets are used to train a neural network-based classifier to identify fault types and locations. A Python based GUI is developed and integrated with the HIL platform to perform real-time fault diagnosis. The trained model is deployed in this environment, and its ability to generalize from synthetic training data to real-time test scenarios is thoroughly evaluated.
\end{itemize}
%%%%%%%%%%%%%%%%%%
%%%%%%%%%%%%%%%% sub B %%%%%%%%%%%%
\subsection{Microgrid Modeling and Data Acquisition}
\label{subsec:microgrid_model}

\begin{figure*}[!t]
    \centering
    \scalebox{0.85}{  % adjust between 0.7–0.95 for desired size
        \includegraphics[width=\textwidth]{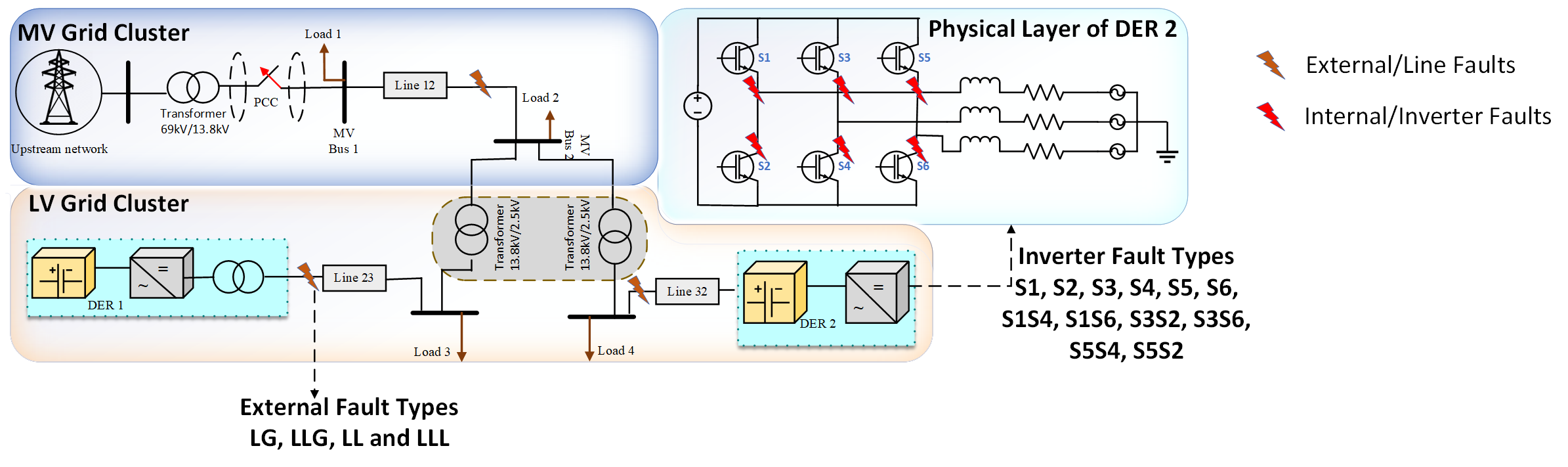}
    }
    \caption{Microgrid architecture and fault classification illustrating the interconnected MV and LV clusters of the microgrid, featuring external faults (LG, LLG, LL and LLL) and internal faults within the inverter's physical layer.}
    \label{Fig2}
\end{figure*}
The experimental framework is implemented within a real-time simulated microgrid environment using the HIL testbed. The system architecture, depicted in Fig.~\ref{Fig2}, comprises two distributed energy resources (DERs) integrated into a hybrid medium-voltage (MV) and low-voltage (LV) cluster configuration. DER~1 corresponds to a photovoltaic (PV) generation unit, while DER~2 represents a battery energy storage system (BESS). These DERs are interconnected via three primary distribution lines: Line~12, Line~23, and Line~32, which respectively represent the Point of Common Coupling (PCC), the BESS interconnection, and the PV feeder. This setup forms the first phase of system development, as illustrated in Fig.~\ref{Fig1}, where datasets are generated under two distinct fault configurations, as detailed below. 
\begin{enumerate}
    \item \textit{External Fault Configurations:} To emulate realistic grid-side faults, three line sections (12, 23, and 32) were subjected to ten different types of short-circuit faults, including single line-to-ground (LG), double line-to-ground (LLG), line-to-line (LL), and three-phase (LLL) conditions. For each fault scenario, variations were introduced in fault resistance ($R_f$), solar irradiance, and load demand to emulate different operating states and improve situational diversity. This process yielded a total of 30 external fault classifiers (10 fault types $\times$ 3 line sections), forming a comprehensive dataset for training and evaluation. The corresponding RMS values of current and voltage were recorded at each line section with eighteen distinct features. This comprehensive feature set served as the foundation for constructing the external fault dataset used in generative modeling and classifier training.
    \item \textit{Internal Fault Configurations:} In addition to external grid faults, internal inverter faults were simulated within the physical layer of DER~2. The inverter bridge consists of six semiconductor switches ($S_1$–$S_6$), and faults were introduced as open-circuit failures by selectively distorting the corresponding PWM control signals.  A total of 12 internal fault classes were generated, comprising six single-switch open-circuit faults and six multiple-switch fault combinations ($S_1S4$, $S_1S6$, $S_3S6$, $S_3S2$, $S_5S2$, and $S_5S4$).  These conditions represent typical hardware degradation or device-level failures that significantly alter current harmonics and voltage profiles. The complete dataset spanning 30 external and 12 internal fault classes forms a rich repository of fault conditions with diverse operational scenarios.  
    %This dual layered dataset enables a comprehensive validation of both statistical diversity and temporal generalization when deploying F2GAN synthesized data for real-time diagnosis.
\end{enumerate}
Deep learning-based generative models have shown significant promise for augmenting high-dimensional tabular datasets. Among these, CGAN, TVAE, and Wasserstein GAN with Gradient Penalty (WGAN-GP) are commonly utilized for synthesizing statistically realistic and structurally coherent samples; hence, these models are considered baseline frameworks \cite{9984938, 10304147}.

%%%%%%%%%%%%%%sub C %%%%%%%% baseline models

\subsection{Conditional Generative Adversarial Network (CGAN)}
\label{subsec:CGAN}
GAN models comprise two competing neural networks: a generator \( G(z, y) \) and a discriminator \( D(x, y) \). The generator aims to synthesize data samples conditioned on class labels \( y \), while the discriminator attempts to distinguish between real samples \( (x, y) \) and generated pairs \( (G(z, y), y) \). Despite the effectiveness of GAN-based frameworks, two major challenges persist: (1) \textit{mode collapse}, where the generator focuses on a limited subset of the data distribution, and (2) \textit{training instability}, as the adversarial optimization struggles to converge to a Nash equilibrium—an ideal point where neither network can improve without altering the other.

The inclusion of label information \( y \) transforms the generation process into a semi-supervised learning paradigm, allowing the model to generate samples for specific fault classes conditionally. This guidance enhances the control and diversity of generated outputs, beneficial for imbalanced or structured datasets. The CGAN training objective follows the standard minimax formulation as in Eq. (\ref{Eq:1}) \cite{10985852}:

\begin{multline}
\min_G \max_D \; 
\mathbb{E}_{x,y \sim p_{\text{real}}}\!\big[\log D(x,y)\big] \\
+ \mathbb{E}_{z \sim p_z, y \sim p_y}\!\big[\log(1 - D(G(z,y),y))\big]
\label{Eq:1}
\end{multline}

\subsection{ Tabular Variational Autoencoder (TVAE)}
\label{subsec:TVAE}
TVAE is a probabilistic generative model designed specifically for heterogeneous tabular data. It operates by encoding input samples into a latent space using an approximate posterior \( q_\phi(z|x) \), and reconstructing the data through a decoder \( p_\theta(x|z) \). The learning objective combines a reconstruction term and a regularization term, expressed as in Eq. (\ref{Eq:2}) \cite{10171153}:

\begin{equation}
\mathcal{L}_{\text{TVAE}} = \mathbb{E}_{q_\phi(z|x)}[\log p_\theta(x|z)] - \text{KL}(q_\phi(z|x) \| p(z)).
\label{Eq:2}
\end{equation}

Here, the Kullback–Leibler (KL) divergence encourages the latent variable \( z \) to follow a prior distribution \( p(z) \), typically chosen as a multivariate standard Gaussian. This regularization prevents overfitting by constraining the geometry of the latent space and promotes generalization in data synthesis. While TVAE is effective in capturing marginal distributions and supports both categorical and continuous attributes, it lacks an adversarial component. Consequently, it may generate overly smooth samples with limited diversity and weaker inter-feature dependencies, which can be suboptimal for datasets with strong physical relationships such as power system fault diagnostics.

\subsection{WGAN-GP}
\label{subsec:WGAN}
The Wasserstein GAN with Gradient Penalty (WGAN-GP) introduces the Wasserstein-1 distance to improve convergence stability and reduce mode collapse. The training objective is expressed as in Eq. (\ref{Eq:3}) \cite{11012665}:
\begin{multline}
\mathcal{L}_{\text{WGAN-GP}} = 
\mathbb{E}_{\tilde{x} \sim p_g}[D(\tilde{x})] 
- \mathbb{E}_{x \sim p_{\text{real}}}[D(x)] \\
+ \lambda \, \mathbb{E}_{\hat{x} \sim p_{\hat{x}}}
\left[\left(\|\nabla_{\hat{x}} D(\hat{x})\|_2 - 1\right)^2\right].
\label{Eq:3}
\end{multline}

This formulation ensures smoother discriminator gradients and more stable training. However, despite its improved fidelity, WGAN-GP does not explicitly preserve feature-wise dependencies crucial for physically consistent fault data.

%%%%%%%%%%%%%%%%%%%%%%

\subsection{Problem Formulation}
\label{sec:problem_formulation}

Let the microgrid fault dataset be represented as in Eq. (\ref{Eq:4})
\begin{equation}
 \mathcal{D} = \{(x_i, y_i)\}_{i=1}^{N} 
 \label{Eq:4}
\end{equation}

where, $x_i \in \mathbb{R}^d$ denotes the $d$-dimensional feature vector
comprising RMS voltage and current magnitudes at multiple line sections, 
and $y_i \in \{1, 2, \ldots, C\}$ represents the corresponding fault class label. 
For the internal fault dataset, $C = 12$ classes, while the external fault dataset contains $C = 30$ distinct fault types. The modeling, dataset formulation, and subsequent generative analysis for internal and external faults were conducted independently, ensuring that each scenario captures its distinct fault dynamics and physical response characteristics without cross-domain overlap.

Due to the limited occurrence of fault events, 
$\mathcal{D}$ is both \textit{imbalanced} and \textit{scarce}, 
which can be expressed as Eq. (\ref{Eq:5}):
\begin{equation}
N_c \ll N_{\text{max}} \quad 
\forall c \in \{1,\ldots, C\}, \quad
N_c = |\{i \mid y_i = c\}|
\label{Eq:5}
\end{equation}

where $N_{max}$ denotes the cardinality of the majority class.  
The goal of synthetic data generation is thus to construct a generative mapping
$G: (z, y) \rightarrow \hat{x}$,
where $z \sim p_z(z)$ is a latent random vector
and $\hat{x}$ is the generated feature vector conditioned on class $y$,
such that the synthetic distribution $p_g(x|y)$ approximates the real distribution $p_r(x|y)$ as in Eq. (\ref{Eq:6}):
\begin{equation}
p_g(x|y) \approx p_r(x|y), \quad
\forall y \in \{1,\ldots, C\}.
\label{Eq:6}
\end{equation}

CGANs approach this by optimizing the adversarial min–max objective as in Eq. (\ref{Eq:7}):
\begin{multline}
\min_G \max_D \;
\mathcal{V}(D,G) =
\mathbb{E}_{x \sim p_r}[\log D(x,y)] \\
+ \mathbb{E}_{z \sim p_z}[\log(1 - D(G(z,y), y))].
\label{Eq:7}
\end{multline}

However, in high-dimensional fault datasets, 
the mapping in Eq.(~\ref{Eq:6}) becomes unstable because the discriminator $D$
primarily focuses on marginal similarities in feature space,
neglecting inter-feature dependencies and physical correlations intrinsic to power system measurements.
Let $\mu_r, \sigma_r^2, \rho_r$ represent the mean, variance, and correlation matrix
of the real features, and $\mu_g, \sigma_g^2, \rho_g$ their generated counterparts.
Then, statistical inconsistency is quantified as in Eq. (\ref{Eq:8}):
\begin{equation}
\Delta_\text{stat} = 
\|\mu_r - \mu_g\|_2^2
+ \|\sigma_r^2 - \sigma_g^2\|_2^2
+ \|\rho_r - \rho_g\|_F^2.
\label{Eq:8}
\end{equation}

In conventional GANs, the generator receives no direct gradient feedback 
from $\Delta_\text{stat}$, leading to mode collapse or physically incoherent data that fails to preserve voltage–current correlations. Therefore, the central problem addressed in this paper is: How to synthesize statistically balanced and physically coherent 
fault datasets that capture inter-feature dependencies and remain transferable from offline domains to real-time temporal fault conditions.

This motivates the introduction of a F2GAN architecture, where $\Delta_\text{stat}$ is explicitly embedded into the generator’s optimization
through differentiable feedback terms.

\subsection{Proposed F2GAN Architecture}
\label{subsec:F2gan}
\begin{figure}[!t]
    \centering
    \includegraphics[width=\columnwidth]{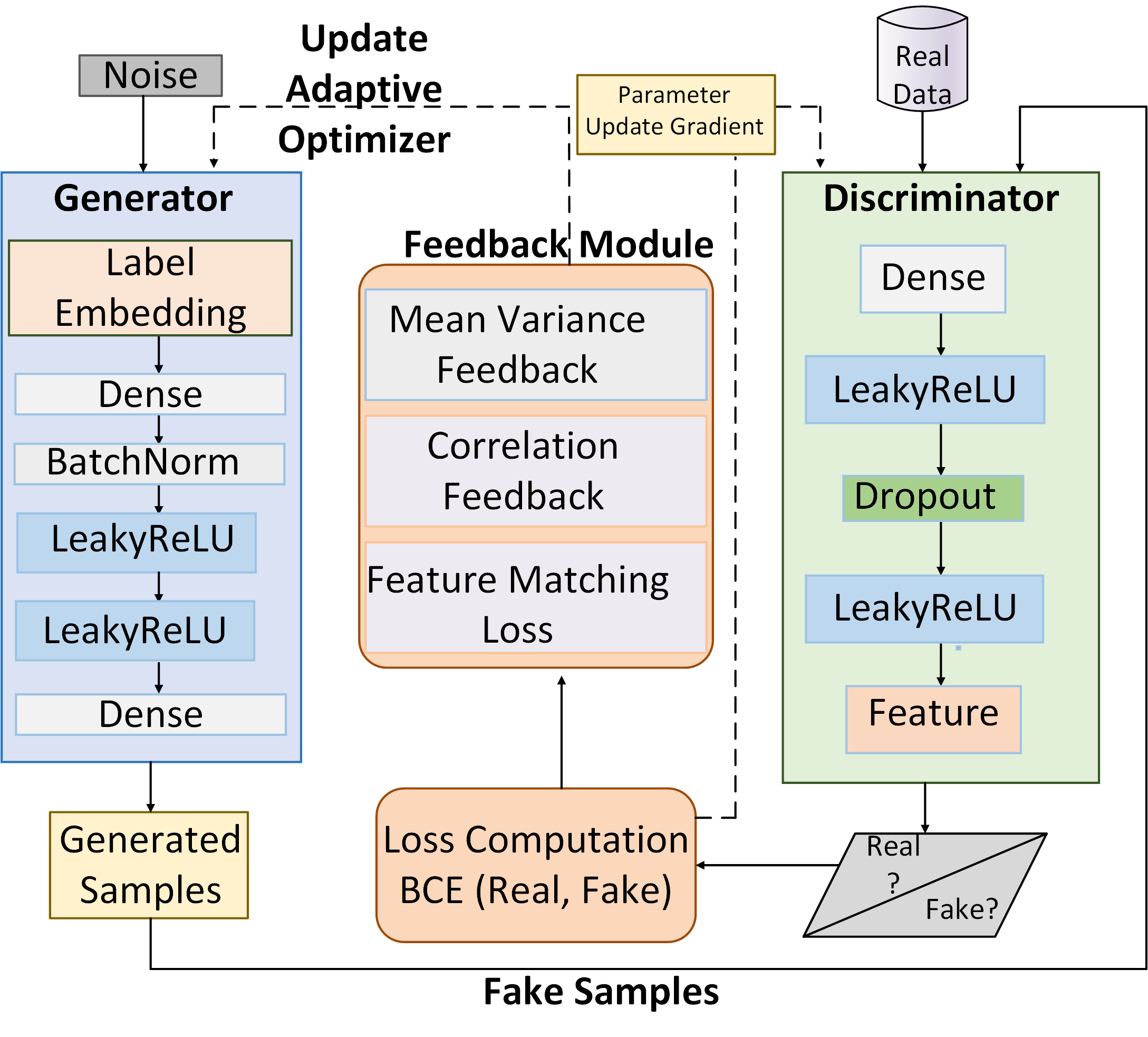}  % Adjusted to fit one column
    \caption{Proposed F2GAN architecture with a feedback module providing mean–variance and correlation feedback to enhance generator learning.}
    \label{Fig3}
\end{figure}
%To address the limitations of conventional GANs that only minimize adversarial divergence
%without preserving the intrinsic statistical structure of power-system fault data,
%a \textbf{Feature-Feedback Generative Adversarial Network (F2GAN)} is proposed.
%The central idea is to provide the generator with explicit statistical feedback derived from
%the discriminator’s internal feature space—capturing mean, variance, and inter-feature correlation—thereby enforcing physical coherence in the synthesized fault patterns.
Figure~\ref{Fig3} provides the overview of the F2GAN training pipeline. The generator receives both noise and encoded labels as input, generating synthetic fault data samples. These fake samples, along with real simulated fault samples, are passed through the discriminator, which outputs both classification logits and deep feature representations. A dedicated feedback module computes the discrepancy between real and synthetic features using three key statistical feedbacks: mean–variance alignment, correlation preservation, and feature matching loss. This feedback is used to refine the generator via an adaptive optimizer, ensuring statistically grounded and physically plausible outputs.

Let \( f_r = f(x) \) and \( f_g = f(G(z)) \) denote the feature embeddings of real and generated samples, respectively, as extracted from the intermediate feature layer of the discriminator.  
Their empirical statistical properties are defined as:
\[
\mu_r = \mathbb{E}[f_r], \quad
\mu_g = \mathbb{E}[f_g], \quad
\sigma_r^2 = \mathbb{V}[f_r], \quad
\sigma_g^2 = \mathbb{V}[f_g],
\]
\[
\rho_r = \text{corrcoef}(f_r^\top), \quad
\rho_g = \text{corrcoef}(f_g^\top),
\]
where \( \mathbb{E}[\cdot] \) and \( \mathbb{V}[\cdot] \) denote the empirical mean and variance, respectively, and \( \text{corrcoef}(\cdot) \) denotes the Pearson correlation matrix computed across feature dimensions. 

The F2GAN generator is trained using a composite loss function that incorporates three distinct feedback components: adversarial loss, mean–variance alignment, and correlation consistency.

\paragraph*{(1) Adversarial Loss}
The generator aims to deceive the discriminator as in the standard GAN formulation:
\begin{equation}
\mathcal{L}_{adv}
= \mathbb{E}_{z \sim p_z} [\, \log(1 - D(G(z))) \,].
\end{equation}

\paragraph*{(2) Mean–Variance Feedback}
This feedback aligns the first- and second-order statistics between real and generated feature distributions:
\begin{equation}
\mathcal{L}_{MV}
= \|\mu_r - \mu_g\|_2^2 + \|\sigma_r - \sigma_g\|_2^2 .
\end{equation}
It ensures that the generator captures both the steady-state level and the dynamic spread of
voltage–current features during fault events.

\paragraph*{(3) Correlation Feedback}
To maintain cross-feature dependencies (e.g., coupling between three-phase voltages and currents),
a correlation-preserving term is defined as:
\begin{equation}
\mathcal{L}_{corr}
= \|\rho_r - \rho_g\|_F^2 ,
\end{equation}
where $\|\cdot\|_F$ denotes the Frobenius norm.

\paragraph*{(4) Final Generator Objective}
Combining the three components, the complete generator loss is expressed as:
\begin{equation}
\mathcal{L}_G
= \mathcal{L}_{adv}
+ \lambda_{MV}\,\mathcal{L}_{MV}
+ \lambda_{corr}\,\mathcal{L}_{corr},
\label{eq:f2gan_loss}
\end{equation}
where $\lambda_{MV}$ and $\lambda_{corr}$ are hyperparameters controlling
the strength of statistical feedback and correlation regularization, respectively.
The discriminator retains the conventional binary cross-entropy objective:
\begin{equation}
\mathcal{L}_D
= - \mathbb{E}_{x \sim p_{real}}[\log D(x)]
  - \mathbb{E}_{z \sim p_z}[\log(1 - D(G(z)))] .
\end{equation}
%%%%%%%%%%%%%%%%%%%%%%%%%%%algorithm
\begin{algorithm}[t]
\caption{F2GAN Training with Statistical Feedback}
\label{alg:f2gan}
\begin{algorithmic}[1]
\Require Labeled dataset $\mathcal{D} = \{(x_i, y_i)\}_{i=1}^{N}$; label encoder; scaler to $[-1,1]$;
\Statex batch size $B$; epochs $T$; noise prior $p_z$; weights $(\lambda_{\text{MV}}, \lambda_{\text{corr}})$;
\Statex label smoothing $\alpha \in (0,1)$; gradient clip $\gamma$.
\Ensure Trained generator $G$ and discriminator $D$.
\vspace{1mm}

\State Encode labels and normalize features.
\State Build mini-batches $(X_r, Y)$ of size $B$.
\State Initialize $G$ (takes $[z, y]$) and $D$ (outputs logits and feature vector $f(\cdot)$).

\For{$t = 1$ to $T$} \Comment{Training epochs}
    \ForAll{mini-batch $(X_r, Y)$}
        \State \textbf{// Discriminator Step}
        \State Sample noise $Z \sim p_z$; generate fake samples \\ 
        \hspace*{3em} $X_f \gets G(Z, Y)$.
        \State Compute logits: $\ell_r \gets  D(X_r, Y)$; \\
        \hspace*{3em} $\ell_f \gets  D(X_f, Y)$.
       \State Compute discriminator loss:\\
        \hspace*{3em} $\mathcal{L}_D \gets \begin{aligned}[t]
        & -\mathbb{E}_{x \sim p_{\text{real}}}[\log D(x)] \\
        & -\mathbb{E}_{z \sim p_z}[\log(1 - D(G(z)))]
        \end{aligned}$

        \State Update $D$ using $\nabla \mathcal{L}_D$; clip gradients $\|\nabla\| \le \gamma$

        \vspace{1mm}

        \State \textbf{// Generator Step (Adversarial + Feedback)}
        \State Sample new noise $Z' \sim p_z$; \\ \hspace*{3em} generate $X'_f \gets G(Z', Y)$.
        \State Obtain features: $(\ell'_f, F_g) \gets D(X'_f, Y)$; \\
       \hspace*{3em}  $(\cdot, F_r) \gets D(X_r, Y)$ \Comment{stop-grad on $F_r$}
        
        \State Compute adversarial loss:
        $\mathcal{L}_{adv} $.
        %$\mathcal{L}_{adv} \gets -\log \sigma(\ell'_f)$.

        \State Compute mean–variance feedback:
        \State \hspace{1em} $\mu_r \gets \text{mean}(F_r)$; $\mu_g \gets \text{mean}(F_g)$
        \State \hspace{1em} $\sigma_r \gets \text{std}(F_r)$; $\sigma_g \gets \text{std}(F_g)$
        \State \hspace{1em} $\mathcal{L}_{MV} \gets \|\mu_r - \mu_g\|_2^2 + \|\sigma_r - \sigma_g\|_2^2$

        \State Compute correlation feedback:
        \State \hspace{1em} $\rho_r \gets \text{corrcoef}(F_r^\top)$; $\rho_g \gets \text{corrcoef}(F_g^\top)$
        \State \hspace{1em} $\mathcal{L}_{corr} \gets \|\rho_r - \rho_g\|_F^2$

        \State Combine losses:
        \State \hspace{1em} $\mathcal{L}_G \gets \mathcal{L}_{adv} + \lambda_{MV} \mathcal{L}_{MV} + \lambda_{corr} \mathcal{L}_{corr}$
        \State Update $G$ using $\nabla \mathcal{L}_G$; clip gradients $\|\nabla\| \le \gamma$
    \EndFor
\EndFor

\vspace{2mm}
\State \textbf{Post-Training Balanced Synthesis}
\ForAll{class $c \in \mathcal{Y}$}
    \State Sample $K$ latent vectors; generate $K$ samples $X_c \gets \{G(z, c)\}$.
    \State Inverse-scale $X_c$ to original feature space.
\EndFor
\State \Return $G$, $D$, and final balanced dataset $\widetilde{\mathcal{D}} = \bigcup_c X_c$
\end{algorithmic}
\end{algorithm}

%%%%%%%%%algorithm
\paragraph*{(5) Algorithmic Workflow}
The entire training strategy is detailed in Algorithm~\ref{alg:f2gan}, which mirrors the pipeline in Fig.~\ref{Fig3}. Each iteration involves:
\begin{itemize}
    \item Forwarding real and fake samples through the discriminator to extract logits and features.
    \item Computing feedback terms ($\mathcal{L}_{MV}$ and $\mathcal{L}_{corr}$) using feature statistics.
    \item Updating the generator using the combined loss function $\mathcal{L}_G$.
    \item To ensure stable adversarial training, we apply gradient clipping with a threshold $\gamma = 0.5$.
\end{itemize}
The feedback module in the Fig.~\ref{Fig3}  corresponds directly to the mean, variance, and correlation computation blocks in Algorithm~\ref{alg:f2gan}.
\paragraph*{Training Strategy}
During training, the generator and discriminator are updated alternately.
The discriminator provides continuous feedback to the generator by comparing
feature-space statistics between real and synthetic samples.
Gradient clipping and label smoothing are employed to enhance convergence stability,
and the feedback terms prevent mode collapse by maintaining statistical diversity.

Overall, this formulation enables F2GAN to learn the underlying
\textit{fault signature manifold} of microgrid events,
bridging statistical fidelity with physical interpretability.

\section{ Experimental Setup and Evaluation Metrics}
\label{sub:Evaluation Metrics}
\subsection{Experimental setup}
First, a microgrid model, as described in Section \ref{subsec:microgrid_model}, was developed and deployed on the HIL testbed. To comprehensively evaluate the proposed model, two distinct datasets were generated: one focusing on external fault analysis and the other on inverter fault detection within a microgrid environment.

Both datasets were obtained through extensive offline simulations conducted under diverse operating conditions to ensure strong model generalization. The external fault dataset comprises 30 fault classes, with each sample consisting of 18 electrical features, specifically the RMS voltage and current values across all three phases, resulting in a total of 6,000 samples.

In contrast, the internal fault dataset captures the failure behavior of the inverter’s switching network, encompassing 12 fault classes that span both single-switch and multi-switch fault conditions. This dataset includes 2,000 samples and was specifically designed to assess the model's robustness in limited-data scenarios, which are often encountered in real-world inverter diagnostics.

Together, these two datasets cover a broad spectrum of operating and fault conditions, including variations in power balance, load dynamics, and mode transitions. This diversity provides a comprehensive basis to evaluate the generalization capability of the proposed F2GAN model across multiple fault scenarios. The F2GAN and all baseline models were implemented in the PyTorch framework. The results of the evaluation are presented using three approaches:
\begin{enumerate}
    \item \textit{Qualitative analysis}: through kernel density estimation (KDE) plots
    \item \textit{Quantitative analysis}: using multiple statistical performance indices.
    \item \textit{Train on Synthetic, Test on Real Evaluation (TSTR)}: assesses the practical utility of synthetic data by training models solely on generated samples and testing them on real data. It verifies whether the synthetic dataset preserves class semantics and decision boundaries, ensuring its effectiveness for real-world fault diagnosis
\end{enumerate}
While KDE plots provide a visual perspective, they are often insufficient for a rigorous assessment, as it is difficult for human observers to judge the statistical similarity between real and synthetic samples. Therefore, quantitative metrics are essential for a thorough and objective evaluation. The mathematical evaluation of these evaluation metrics is explained in the following section.

\subsection{Evaluation Metrics}

Three statistical distance measures were employed to evaluate the fidelity between
the real and generated datasets: they are Wasserstein distance ($\mathcal{W}$),
the Maximum Mean Discrepancy (MMD), and the Kolmogorov--Smirnov (KS) statistic.
Each metric captures complementary aspects of similarity between the real
distribution $P_r$ and the generated distribution $P_g$.

\paragraph*{1) Wasserstein Distance}
The Wasserstein distance quantifies the optimal transport cost between two probability
distributions as given by Eq.(~\ref{Eq:15}) \cite{10208147}:
\begin{equation}
\mathcal{W}(P_r, P_g) = 
\inf_{\gamma \in \Pi(P_r, P_g)} 
\mathbb{E}_{(x,y)\sim\gamma}[\|x - y\|],
\label{Eq:15}
\end{equation}
where $\mathcal{W}(P_r, P_g)$ denotes the Wasserstein (Earth Mover’s) distance between 
the real data distribution $P_r$ and the generated data distribution $P_g$.  
Here, $\gamma \in \Pi(P_r, P_g)$ defines a joint coupling distribution over the pairs 
$(x, y)$ such that its marginals are $P_r$ and $P_g$, respectively; 
$\mathbb{E}_{(x,y)\sim\gamma}[\cdot]$ denotes the expectation over this coupling; 
and $\|x - y\|$ represents the $L_2$ distance between a real sample $x$ and a 
synthetic sample $y$.  
This formulation captures the minimal expected cost required to transform one 
distribution into another, providing a stable and geometry-aware measure of generative fidelity.

\paragraph*{2) Maximum Mean Discrepancy (MMD)}
The MMD evaluates the difference between the mean embeddings of $P_r$ and $P_g$ in
a reproducing kernel Hilbert space (RKHS), capturing mean and variance alignment as in Eq. (\ref{Eq:16}):
\begin{align}
\mathrm{MMD}^2(P_r, P_g) &=
\mathbb{E}_{x,x'\sim P_r}[k(x,x')] 
+ \mathbb{E}_{y,y'\sim P_g}[k(y,y')] \nonumber\\
&\quad - 2\,\mathbb{E}_{x\sim P_r,\,y\sim P_g}[k(x,y)],
\label{Eq:16}
\end{align}
where $k(\cdot,\cdot)$ is a Gaussian kernel defined as in in \ref{Eq:17}
\begin{equation}
k(x,y)=\exp\!\left(-\frac{\|x-y\|^2}{\sigma^2}\right).
\label{Eq:17}
\end{equation}

\paragraph*{3) Kolmogorov--Smirnov Statistic}
The KS statistic measures the maximum deviation between the empirical cumulative
distribution functions (CDFs) of $P_r$ and $P_g$ as in \ref{Eq:18}: 
\begin{equation}
\mathrm{KS}(P_r,P_g)=
\sup_x \big|F_r(x)-F_g(x)\big|,
\label{Eq:18}
\end{equation}
where $F_r$ and $F_g$ denote the empirical CDFs of the real and generated data,
respectively.

All three metrics were computed feature-wise and averaged across the dataset to
provide a global measure of statistical similarity between real and synthetic fault data.
\paragraph*{4) Train-on-Synthetic, Test-on-Real (TSTR) Evaluation}

To assess the usability and generalization capability of the generated synthetic data, a TSTR evaluation framework was employed. In this approach, multiple classifiers, such as Support Vector Machine (SVM), Decision Tree (DT), K-Nearest Neighbors (KNN), and Neural Network (NN), were trained exclusively on synthetic samples produced by each generative model and subsequently tested on the corresponding real datasets. The resulting accuracy, precision, recall, and F1-score metrics provide a quantitative measure of how effectively the synthetic data captures the underlying physical and statistical characteristics of the real system. Higher TSTR performance indicates that the generated data retains discriminative information necessary for reliable fault diagnosis.

\section{Results and Discussion}
\label{sec:Results}
\subsection{Qualitative Analysis}
\begin{figure}[!t]
    \centering
    \includegraphics[width=\columnwidth, height=0.9\textheight, keepaspectratio]{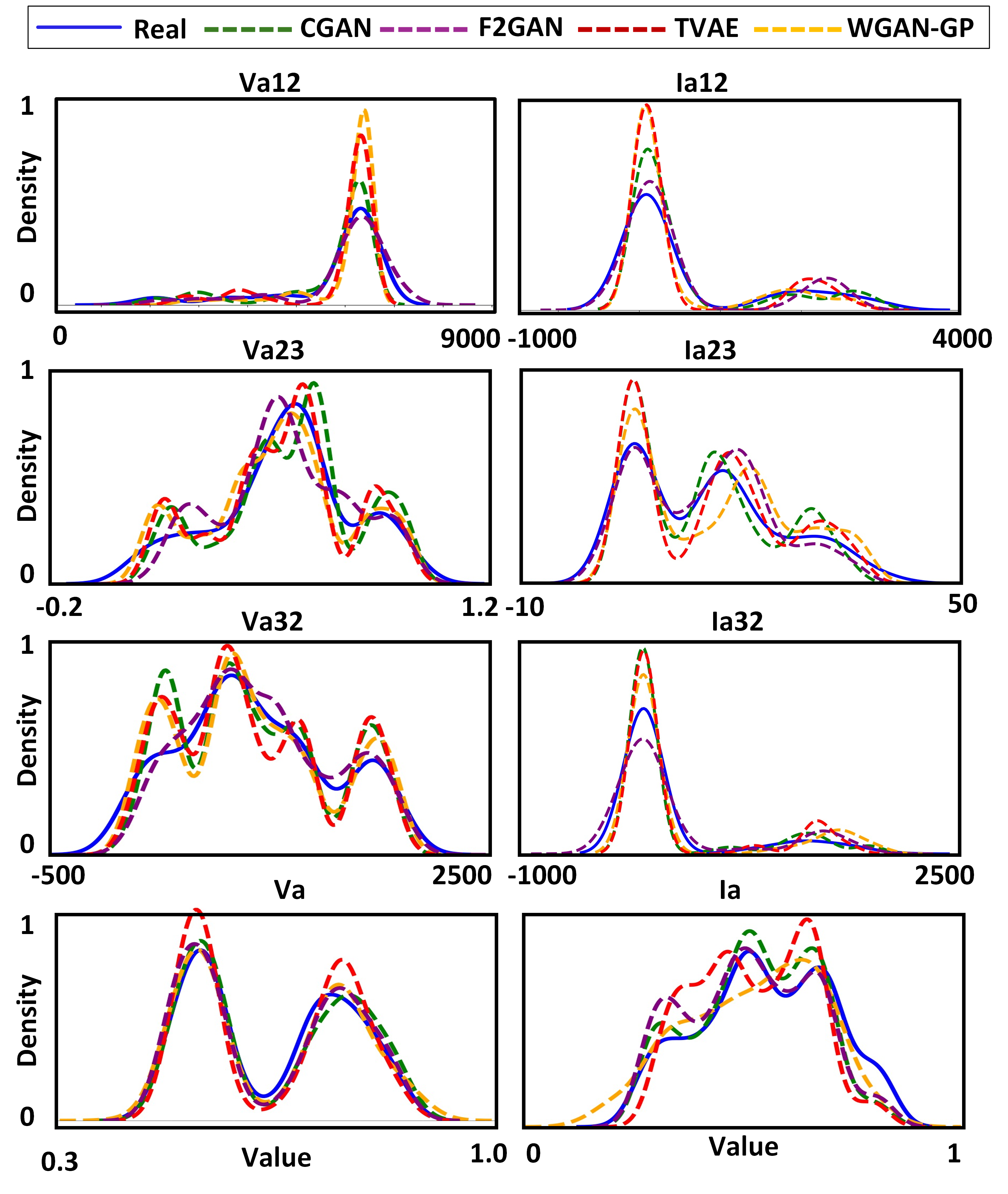}
    \caption{KDE plots comparing real and synthetic distributions for voltage–current pairs in phase-a line sections 12 and 23 under external faults. F2GAN displays strong density alignment with real data.}
    \label{Fig4}
\end{figure}

The KDE plots in Fig. \ref{Fig4}  visualize the statistical density distributions of voltage and current RMS features for phase A across each of the three line sections: Line 12, Line 23, and Line 32. Each figure corresponds to the phase A voltage and current of every line section. These plots help assess the fidelity of the generated samples by comparing their density profiles with those of real simulated data. A high degree of overlap in peak position, spread, and contour shape between real and synthetic KDE curves signifies strong distributional alignment.

Across all line sections and fault conditions, F2GAN exhibits significantly better alignment with the simulated (Real) data than baseline models, such as CGAN, WGAN-GP, and TVAE. Specifically, the synthetic curves from F2GAN closely follow the sharpness and smoothness of the real distributions, indicating that both amplitude fidelity and inter-feature relationships are well preserved. This behavior is particularly prominent in external fault plots, where phase current spikes are realistically captured, and in internal inverter faults, where subtler shifts are also replicated with minimal mode collapse. These results collectively demonstrate F2GAN’s ability to maintain statistical consistency across fault categories.

\subsection{Quantitative analysis}
The quantitative evaluation comprises two complementary validation strategies: statistical fidelity analysis and task-based generalization. Table~\ref{tab:dist_fidelity} reports the \textit{distributional similarity metrics}: Wasserstein Distance, MMD, and KS statistics computed between real and synthetic datasets for both internal and external fault scenarios. These metrics reflect how well the synthetic data preserve the underlying probability distributions of the real data, where lower values indicate better fidelity. The proposed F2GAN consistently achieves the lowest distances across all metrics, with values of 307.09 (external) and 1.2899 (internal) for the Wasserstein distance, demonstrating its superior ability to replicate realistic feature distributions compared to CGAN, WGAN-GP, and CVAE.

\begin{figure*}[!t]
\centering
\includegraphics[width=1.0\linewidth]{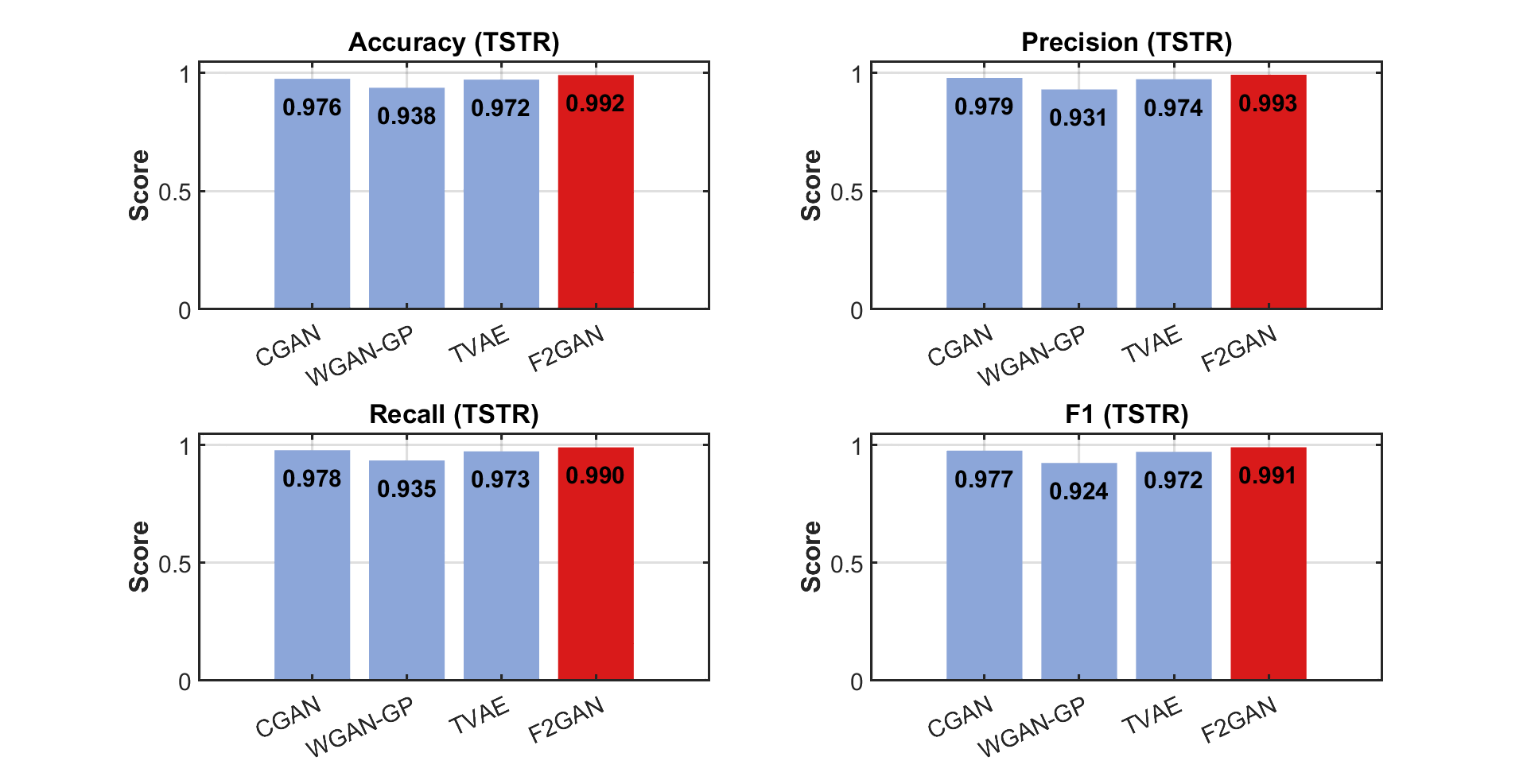}   
\caption{TSTR evaluation metrics across generative models (CGAN, WGAN-GP, TVAE, F2GAN). }
\label{fig:tstr_results}
\end{figure*}

Fig.~\ref{fig:tstr_results} presents a comprehensive comparison of four synthetic data generation models: CGAN, WGAN-GP, TVAE and the proposed F2GAN evaluated using the TSTR framework across four key classification metrics: Accuracy, Precision, Recall, and F1-score. These metrics quantify the extent to which classifiers trained on synthetic data can generalize to real-world fault scenarios.

Among all models, F2GAN consistently outperforms the others, achieving the highest performance across all metrics: an accuracy of 0.992, Precision of 0.993, Recall of 0.990, and F1-score of 0.991. In comparison, CGAN also shows strong performance with values of 0.976 (Accuracy), 0.979 (Precision), 0.978 (Recall), and 0.977 (F1), but slightly trails behind F2GAN. Meanwhile, WGAN-GP displays slightly low values of Accuracy: 0.938, F1: 0.924, indicating less reliable generalization. CVAE performs moderately well (Accuracy: 0.972, F1: 0.972), but again falls short of F2GAN.

These results confirm that F2GAN not only preserves the statistical fidelity of the original dataset but also enables effective downstream classification in real-time diagnostic pipelines, reinforcing its suitability for high-fidelity synthetic data generation in power system fault analysis.

To provide further granularity, Tables~\ref{tab:TSTR_internal} and~\ref{tab:TSTR_external} detail the \textit{TSTR classification performance} for internal and external fault datasets, respectively, using Decision Tree (DT), K-Nearest Neighbors (KNN), Neural Network (NN), and Support Vector Machine (SVM) classifiers. F2GAN consistently achieves the highest per-model accuracy scores, leading to the best average TSTR performance: 0.992 for internal faults and 0.990 for external faults. These results confirm that F2GAN not only preserves statistical properties but also captures critical discriminative features, making it highly effective for real-time fault classification tasks in microgrids.

\begin{table}[htbp]
\centering
\caption{Distributional Fidelity Metrics Comparison for Internal and External Fault Datasets}
\label{tab:dist_fidelity}
\renewcommand{\arraystretch}{1.15}
\setlength{\tabcolsep}{5pt}
\begin{tabular}{lcccc}
\hline
\textbf{Dataset} & \textbf{Model} & \textbf{Wasserstein} $\downarrow$ & \textbf{MMD} $\downarrow$ & \textbf{KS} $\downarrow$ \\
\hline
\multirow{4}{*}{\textbf{External}} 
& CGAN             & 459.16  & 0.00776 & 0.5787 \\
& WGAN-GP          & 365.81  & 0.00776 & 0.3919 \\
& CVAE             & 455.16  & 0.00779 & 0.6463 \\
& \textbf{Proposed F2GAN} & \textbf{307.09}  & \textbf{0.00776} & \textbf{0.3434} \\
\hline
\multirow{4}{*}{\textbf{Internal}} 
& CGAN             & 1.3921  & 0.01946 & 0.1219 \\
& WGAN-GP          & 1.5301  & 0.01935 & 0.1342 \\
& CVAE             & 1.5807  & 0.01622 & 0.1395 \\
& \textbf{Proposed F2GAN} & \textbf{1.2899}  & \textbf{0.01622} & \textbf{0.1310} \\
\hline
\end{tabular}
\end{table}

\begin{table}[h!]
\centering
\caption{TSTR Evaluation across Classifiers for Internal Fault Dataset}
\label{tab:TSTR_internal}
\renewcommand{\arraystretch}{1.15}
\begin{tabular}{lcccc}
\hline
\textbf{Classifier} & \textbf{CGAN} & \textbf{WGAN-GP} & \textbf{CVAE} & \textbf{Proposed F2GAN} \\
\hline
Decision Tree & 0.902 & 0.982 & 0.919 & \textbf{0.983} \\
KNN & 0.980 & 0.965 & 0.995 & \textbf{0.993} \\
Neural Net & 0.997 & 0.980 & 0.998 & \textbf{0.997} \\
SVM & 0.996 & 0.979 & 0.997 & \textbf{0.996}\\
\hline
\textbf{Average} & 0.969 & 0.977 & 0.978 & \textbf{0.992} \\
\hline
\end{tabular}
\end{table}

\begin{table}[h!]
\centering
\caption{TSTR Evaluation across Classifiers for External Fault Dataset}
\label{tab:TSTR_external}
\renewcommand{\arraystretch}{1.15}
\begin{tabular}{lcccc}
\hline
\textbf{Classifier} & \textbf{CGAN} & \textbf{WGAN-GP} & \textbf{CVAE} & \textbf{Proposed F2GAN} \\
\hline
Decision Tree & 0.929 & 0.815 & 0.865 & \textbf{0.963} \\
KNN & 0.999 & 0.919 & 0.998 & \textbf{1.000} \\
Neural Net & 0.998 & 0.913 & 0.999 & \textbf{1.000} \\
SVM & 0.998 & 0.926 & 0.998 & \textbf{1.000} \\
\hline
\textbf{Average} & 0.981 & 0.893 & 0.965 & \textbf{0.990} \\
\hline
\end{tabular}
\end{table}
\subsection{Validation on Real-Time Testbed}
\begin{figure}[htbp]
    \centering

    % -------- First subfigure --------
    \subfloat[]{%
        \includegraphics[width=0.8\linewidth]{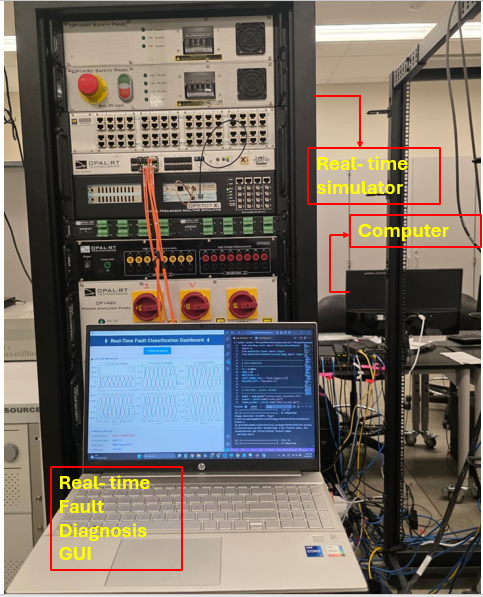}%
        \label{fig:hil_setup}
    }\\[1em] % vertical space between subfigures

    % -------- Second subfigure --------
    \subfloat[]{%
        \includegraphics[width=0.8\linewidth]{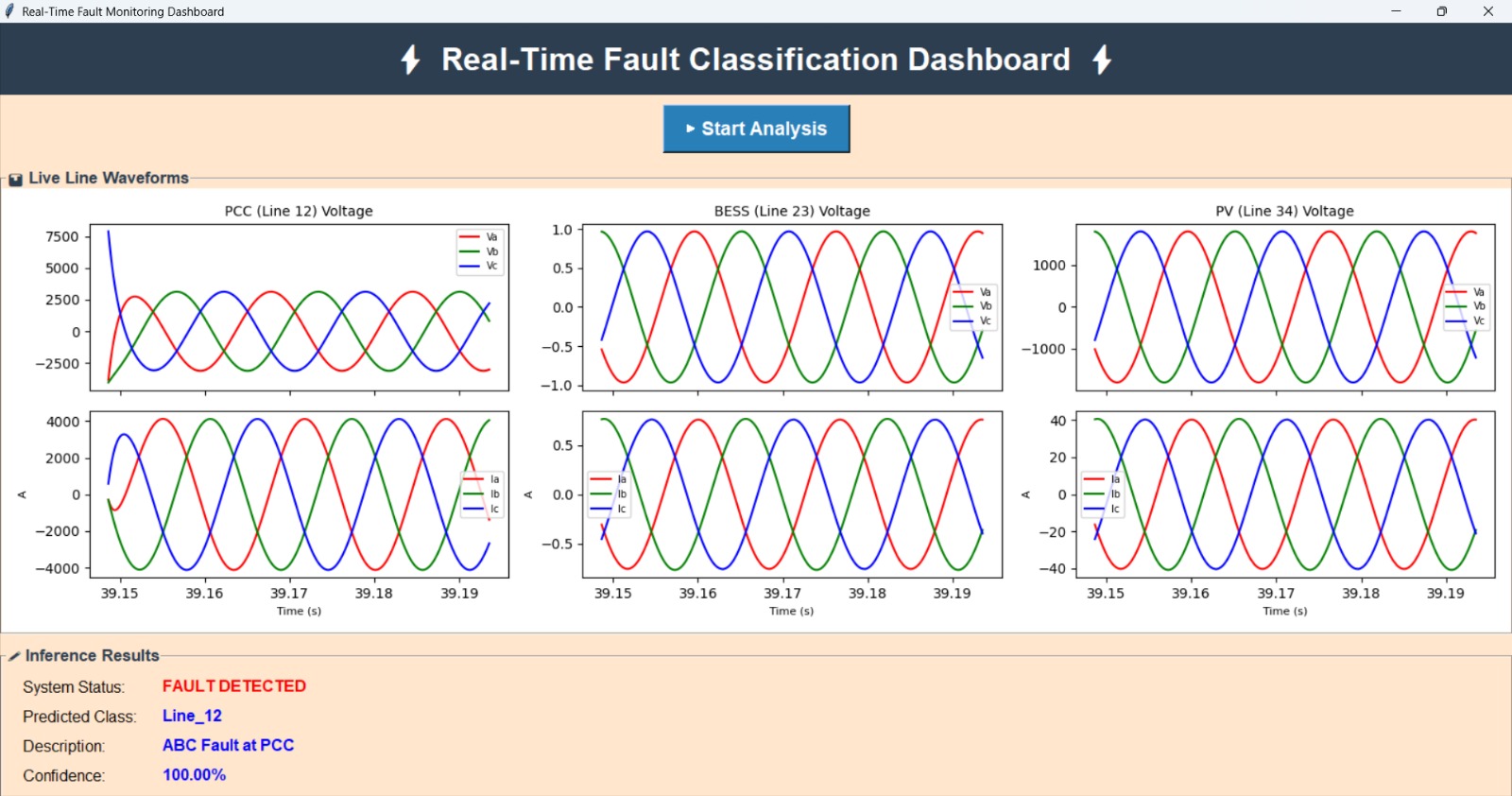}%
        \label{fig:hil_dashboard}
    }

    \caption{Real-time hardware-in-the-loop testbed with monitoring system. 
    The simulator executes fault scenarios and streams measurement data to the diagnosis GUI, 
    where the trained ML model classifies fault type and location in real time.}
    \label{HIL}
\end{figure}

To validate the real-time applicability and robustness of the proposed framework, a HIL test environment was developed, as illustrated in Fig.~\ref{fig:hil_setup}. The microgrid model runs continuously on the simulator with programmable external fault scenarios that can be triggered dynamically during operation. The simulator is interfaced with a Python-based GUI as illustrated in Fig.~\ref{fig:hil_dashboard}, which integrates both the data preprocessing pipeline and the trained neural network classifier. The neural network model was trained exclusively on synthetic fault data generated by the proposed F2GAN architecture.

During each experiment, when a fault event is initiated on the real-time simulator, one complete cycle of data including phase voltages, currents, frequency, and phase angles are captured and stored in a dedicated fault data file along with a timestamp. The GUI continuously monitors this file and retrieves the most recent waveform data, and converts the raw measurements into RMS representations. These processed values are then passed to the trained classifier for inference.

The system performs instantaneous fault type and location identification upon data acquisition, with results displayed live on the GUI dashboard. Across all test cases, the model achieved 100$\%$ classification accuracy and confidence, successfully detecting and localizing each external fault condition in real-time. This demonstration confirms that the F2GAN-generated synthetic data effectively generalizes to physical measurements, validating the framework’s ability to support real-time fault diagnosis in inverter-dominated microgrids.

\section{Conclusion and Future Scope}

\label{sec:Conclusion}
This paper presents a Feature-Feedback Generative Adversarial Network (F2GAN) framework for generating realistic fault data and validating diagnostics in inverter-dominated microgrids. Two comprehensive datasets were utilized: one for external faults and another for inverter faults to evaluate the model’s performance under diverse operating and fault conditions. Quantitative analyses using statistical distance metrics (MMD, WD, and KS) and TSTR validation confirmed that the proposed F2GAN achieves superior distributional alignment and generalization compared to CGAN, WGAN-GP, and TVAE baselines. Qualitative assessments through KDE plots further demonstrated strong structural resemblance between real and synthetic data.

The trained ML models, when deployed on a HIL testbed, successfully identified external fault types and locations with a 100$\%$ accuracy, thereby validating the transferability of F2GAN-generated synthetic data to practical environments. These findings highlight the effectiveness of synthetic data-driven training in enhancing fault diagnosis reliability without the need for extensive real-world data collection.

Future work will focus on extending the real-time validation framework to cover internal inverter fault scenarios and exploring adaptive domain-transfer mechanisms to enhance robustness across different microgrid configurations. Additionally, integrating the proposed synthetic data generation pipeline with online learning and cyber–physical security layers could further advance the development of resilient and self-learning microgrid fault diagnostic systems.
%%%%%%%%%%%%%%%%%%%%%%

%%%%%%%%%%%%%%%%%%%%%%

\section{Acknowledgment}
This work was supported by Korea Institute of Energy Technology Evaluation and
Planning(KETEP) grant funded by the Korea government(MOTIE)(RS-2023-00231702, Development of an open-integrated platform for distributed renewable energy systems).

\bibliographystyle{IEEEtran}
%\bibliography{IEEEabrv,ref}

\bibliography{ref}

@INPROCEEDINGS{10563400,
  author={Talha, Muhammad and Liang, Xiaodong and Pannell, Jason and Bowes, Austin and Su, Hanguang},
  booktitle={2024 IEEE/IAS 60th Industrial and Commercial Power Systems Technical Conference (I$\&$CPS)}, 
  title={Experimental Study of Open-Switch Faults for Interfacing Inverters in Microgrids Using OPAL-RT Real-time Simulator}, 
  year={2024},
  volume={},
  number={},
  pages={1-8},
  doi={10.1109/ICPS60943.2024.10563400}}

@INPROCEEDINGS{11204575,
  author={Kasimalla, Swetha Rani and Park, Kuchan and Hong, Junho and Kim, Young-Jin and Lee, HyoJong},
  booktitle={2025 IEEE International Conference on Communications, Control, and Computing Technologies for Smart Grids (SmartGridComm)}, 
  title={Deep Learning-Enabled System Diagnosis in Microgrids: A Feature-Feedback GAN Approach}, 
  year={2025},
  volume={},
  number={},
  pages={1-6},
  doi={10.1109/SmartGridComm65349.2025.11204575}}

@ARTICLE{7467562,
  author={Zarei, Seyed Fariborz and Parniani, M.},
  journal={IEEE Transactions on Power Delivery}, 
  title={A Comprehensive Digital Protection Scheme for Low-Voltage Microgrids with Inverter-Based and Conventional Distributed Generations}, 
  year={2017},
  volume={32},
  number={1},
  pages={441-452},
  doi={10.1109/TPWRD.2016.2566264}}

@ARTICLE{6308748,
  author={Ustun, Taha Selim and Ozansoy, Cagil and Ustun, Aladin},
  journal={IEEE Transactions on Power Systems}, 
  title={Fault current coefficient and time delay assignment for microgrid protection system with central protection unit}, 
  year={2013},
  volume={28},
  number={2},
  pages={598-606},
  doi={10.1109/TPWRS.2012.2214489}}

@ARTICLE{10416843,
  author={Zaben, Muiz M. and Worku, Muhammed Y. and Hassan, Mohamed A. and Abido, Mohammad A.},
  journal={IEEE Access}, 
  title={Machine Learning Methods for Fault Diagnosis in AC Microgrids: A Systematic Review}, 
  year={2024},
  volume={12},
  number={},
  pages={20260-20298},
  doi={10.1109/ACCESS.2024.3360330}}

@ARTICLE{10681285,
  author={Fang, Jiashu and Zheng, Le and Liu, Chongru and Su, Chenbo},
  journal={IEEE Transactions on Industrial Informatics}, 
  title={A Data-Driven Case Generation Model for Transient Stability Assessment Using Generative Adversarial Networks}, 
  year={2024},
  volume={20},
  number={12},
  pages={14391-14400},
  doi={10.1109/TII.2024.3452211}}

@ARTICLE{9911768,
  author={Yan, Rong and Yuan, Yuxuan and Wang, Zhaoyu and Geng, Guangchao and Jiang, Quanyuan},
  journal={IEEE Transactions on Power Systems}, 
  title={Active Distribution System Synthesis via Unbalanced Graph Generative Adversarial Network}, 
  year={2023},
  volume={38},
  number={5},
  pages={4293-4307},
  doi={10.1109/TPWRS.2022.3212029}}

@ARTICLE{10609361,
  author={Ye, Wenjie and Yang, Dongmei and Tang, Chenghong and Wang, Wei and Liu, Gang},
  journal={IEEE Access}, 
  title={Combined Prediction of Wind Power in Extreme Weather Based on Time Series Adversarial Generation Networks}, 
  year={2024},
  volume={12},
  number={},
  pages={102660-102669},
  doi={10.1109/ACCESS.2024.3433496}}

@ARTICLE{9765318,
  author={Yuan, Ran and Wang, Bo and Sun, Yeqi and Song, Xuanning and Watada, Junzo},
  journal={IEEE Transactions on Power Systems}, 
  title={Conditional Style-Based Generative Adversarial Networks for Renewable Scenario Generation}, 
  year={2023},
  volume={38},
  number={2},
  pages={1281-1296},
  doi={10.1109/TPWRS.2022.3170992}}

@ARTICLE{10430431,
  author={Lan, Jian and Zhou, Yanzhen and Guo, Qinglai and Sun, Hongbin},
  journal={IEEE Transactions on Power Systems}, 
  title={Data Augmentation for Data-Driven Methods in Power System Operation: A Novel Framework Using Improved GAN and Transfer Learning}, 
  year={2024},
  volume={39},
  number={5},
  pages={6399-6411},
  doi={10.1109/TPWRS.2024.3364166}}

@ARTICLE{11049025,
  author={Lin, Nan and Palensky, Peter and Vergara, Pedro P.},
  journal={IEEE Transactions on Smart Grid}, 
  title={EnergyDiff: Universal Time-Series Energy Data Generation Using Diffusion Models}, 
  year={2025},
  volume={16},
  number={5},
  pages={4252-4265},
  doi={10.1109/TSG.2025.3581472}}

@ARTICLE{11012665,
  author={Wang, Hongzhen and Qin, Boyu and Hong, Shidong and Xu, Xi and Su, Yiwei and Lu, Tingxiang and Ding, Tao},
  journal={IEEE Transactions on Smart Grid}, 
  title={Enhanced GAN-Based Joint Wind-Solar-Load Scenario Generation With Extreme Weather Labelling}, 
  year={2025},
  volume={16},
  number={5},
  pages={4213-4224},
  doi={10.1109/TSG.2025.3573166}}

@ARTICLE{10818412,
  author={Siddiqui, Hafeez Ur Rehman and Brown, Robert and Ali Saleem, Adil and Amjad Raza, Muhammad and Dudley, Sandra},
  journal={IEEE Access}, 
  title={Enhancing Power Grid Reliability With Machine Learning and Auxiliary Classifier Generative Adversarial Networks: A Study on Fault Detection Using the Georgia Electric System Load Dataset}, 
  year={2025},
  volume={13},
  number={},
  pages={2463-2473},
  doi={10.1109/ACCESS.2024.3524061}}

@ARTICLE{10208147,
  author={Hu, Yi and Li, Yiyan and Song, Lidong and Lee, Han Pyo and Rehm, P. J. and Makdad, Matthew and Miller, Edmond and Lu, Ning},
  journal={IEEE Transactions on Smart Grid}, 
  title={MultiLoad-GAN: A GAN-Based Synthetic Load Group Generation Method Considering Spatial-Temporal Correlations}, 
  year={2024},
  volume={15},
  number={2},
  pages={2309-2320},
  doi={10.1109/TSG.2023.3302192}}

@ARTICLE{9854145,
  author={Zhu, Guangya and Zhou, Kai and Lu, Lu and Fu, Yao and Liu, Zhaogui and Yang, Xiaomin},
  journal={IEEE Transactions on Industrial Informatics}, 
  title={Partial Discharge Data Augmentation Based on Improved Wasserstein Generative Adversarial Network With Gradient Penalty}, 
  year={2023},
  volume={19},
  number={5},
  pages={6565-6575},
  doi={10.1109/TII.2022.3197839}}

@ARTICLE{9731506,
  author={Song, Lidong and Li, Yiyan and Lu, Ning},
  journal={IEEE Transactions on Smart Grid}, 
  title={ProfileSR-GAN: A GAN Based Super-Resolution Method for Generating High-Resolution Load Profiles}, 
  year={2022},
  volume={13},
  number={4},
  pages={3278-3289},
  doi={10.1109/TSG.2022.3158235}}

@ARTICLE{10171153,
  author={Razghandi, Mina and Zhou, Hao and Erol-Kantarci, Melike and Turgut, Damla},
  journal={IEEE Transactions on Smart Grid}, 
  title={Smart Home Energy Management: VAE-GAN Synthetic Dataset Generator and Q-Learning}, 
  year={2024},
  volume={15},
  number={2},
  pages={1562-1573},
  doi={10.1109/TSG.2023.3288824}}

@ARTICLE{9984938,
  author={Chundawat, Vikram S and Tarun, Ayush K and Mandal, Murari and Lahoti, Mukund and Narang, Pratik},
  journal={IEEE Transactions on Artificial Intelligence}, 
  title={A Universal Metric for Robust Evaluation of Synthetic Tabular Data}, 
  year={2024},
  volume={5},
  number={1},
  pages={300-309},
  doi={10.1109/TAI.2022.3229289}}

@ARTICLE{9780384,
  author={Hatata, Ahmed Y. and Essa, Mohamed A. and Sedhom, Bishoy E.},
  journal={IEEE Access}, 
  title={Adaptive Protection Scheme for FREEDM Microgrid Based on Convolutional Neural Network and Gorilla Troops Optimization Technique}, 
  year={2022},
  volume={10},
  number={},
  pages={55583-55601},
  doi={10.1109/ACCESS.2022.3177544}}

@INPROCEEDINGS{10116186,
  author={Peter, Nirma and Gupta, Pankaj and Goel, Nidhi},
  booktitle={2023 10th International Conference on Signal Processing and Integrated Networks (SPIN)}, 
  title={Fault Detection and Identification of Fault location in Hybrid Microgrid using Artificial Neural Network}, 
  year={2023},
  volume={},
  number={},
  pages={686-691},
  doi={10.1109/SPIN57001.2023.10116186}}

@ARTICLE{10985852,
  author={Zhang, Wenkang and Hao, Ying and Luo, Shuyu and Li, Kaidi and Wu, Xun and Jin, Zhanpeng},
  journal={IEEE Transactions on Instrumentation and Measurement}, 
  title={Enhanced Open-Circuit Fault Diagnosis in T-Type Inverters Using Conditional Virtual Sample Generation}, 
  year={2025},
  volume={74},
  number={},
  pages={1-10},
  doi={10.1109/TIM.2025.3566852}}

@ARTICLE{10304147,
  author={Abraham, Asha and Mohideen, Habeeb Shaik and Kayalvizhi, R.},
  journal={IEEE Access}, 
  title={A Tabular Variational Auto Encoder-Based Hybrid Model for Imbalanced Data Classification With Feature Selection}, 
  year={2023},
  volume={11},
  number={},
  pages={122760-122771},
  doi={10.1109/ACCESS.2023.3329139}}

\end{document}